\documentclass{elsart}

\usepackage{amsmath}
\usepackage{amssymb}
\usepackage{graphicx}
\usepackage{dcolumn}
\usepackage{bm}

\def\be{\begin{eqnarray}}
\def\ee{\end{eqnarray}}
\def\ba{\begin{array}}
\def\ea{\end{array}}

\begin{document}

\begin{frontmatter}

\title{Theoretical investigation of the electron velocity in quantum Hall
bars, in the out of linear response regime}

\author[l1]{A. Siddiki},
\author[l2]{D. Eksi}
\ead{denizeksi@gmail.com}
\author[l2]{E. Cicek},
\author[l2]{A. I. Mese},
\author[l2]{S. Aktas} and
\author[l3]{T. Hakio\u{g}lu}

\address[l1]{Physics Department, Arnold Sommerfeld Center for Theoretical Physics, and Center for NanoScience
Ludwig-Maximilians-Universit\"at M\"unchen, D-80333 Munich,
Germany}
\address[l2]{Trakya University, Department of Physics, 22030 Edirne, Turkey}
\address[l3]{Department of Physics and UNAM Material
Science and Nanotecnology Research Institute Bilkent University,
Ankara, 06800 Turkey}

\begin{abstract}
We report on our theoretical investigation of the electron velocity
in (narrow) quantum-Hall systems, considering the
out-of-linear-response regime. The electrostatic properties of the
electron system are obtained by the Thomas-Fermi-Poisson nonlinear
screening theory. The electron velocity distribution as a function
of the lateral coordinate is obtained from the slope of the screened
potential within the incompressible strips (ISs). The asymmetry
induced by the imposed current on the ISs is investigated, as a
function of the current intensity and impurity concentration. We
find that the width of the IS on one side of the sample increases
linearly with the intensity of the applied current and decreases
with the impurity concentration.
\end{abstract}
\end{frontmatter}
%

In the conventional models of the quantum Hall effect (QHE), the
Coulomb interaction is ignored and either localization or the 1D
edge states (ESs) was accepted as the explanation. However, recent
experimental~\cite{Ahlswede02:165,Yacoby04:328} and
theoretical~\cite{Guven03:115327,siddiki2004,TobiasK06:h,Romer07}
investigations of the two dimensional electron systems (2DESs)
under strong perpendicular magnetic fields $B$ provided
information about the local quantities, such as potential,
compressibility, current and electron density. The findings point
out clearly the importance of the involved interactions at narrow
($\lesssim 10$ $\mu$m) samples, which manifest themselves by the
formation of the compressible and incompressible regions. More
recently, the experiments performed in the integer QHE
regime~\cite{Neder06:016804} promote the possibility of inferring
interaction mechanisms between the the ESs, using an electronic
version of Mach-Zehnder interferometer. The surprising results, up
to now, cannot be explained within the naive single particle
pictures, where the group velocity of the electrons ($v_{\rm el}$)
is assumed to be constant and the imposed current is believed to
be carried by the B\"uttiker type ESs.

Recently, we have investigated $v_{\rm el}$ depending on the
sample properties, including the Coulomb
interaction~\cite{deniz06}. We used the screening model to
calculate the electron and potential distribution and obtained
$v_{\rm el}$ from the slope of the screened potential at the Fermi
level and also across the ISs, where the current flows from. The
calculations were done at equilibrium and within the linear
response regime. We found that $v_{\rm el}$ strongly depends on
the sample properties, in the case of small currents. Here, we
extend our investigation to a regime, where the imposed external
current $I$ is high enough to change both the electron and
potential distribution, i.e. out-of-linear-response (OLR) regime.
The effect of the current is included self-consistently by
employing the scheme developed by K. G\"uven and R.R.
Gerhardts~\cite{Guven03:115327}.

Here, we confine ourselves to the historical Chklovskii model
geometry~\cite{Chklovskii93:12605}. We assume that electrons and
donors are on the same $xy$ plane ($z=0$) and donors are
distributed homogeneously in the interval $-d<x<d$, where $2d$ is
the sample width. On the other hand electrons are depleted from
the edges and the translational invariant electron channel is
formed in the interval $-b<x<b$, where $|d|>|b|$ and $|d-b|$ is
called the depletion length. We consider spinless electrons by
setting the spin degeneracy to two, $g_s=2$, therefore ISs will
assume only even local filling factors, $\nu(x)$. First we
calculate the electron density and the screened potential from the
following self-consistent (SC) equations: \be n_{\rm el}(x)=\int
dE\,\frac{D(E)}{e^{[E+V(x)-\mu^\star]/k_{\rm{B}}T)}+1}
\label{tfaed} \ee and \be V(x)= - \frac{2e^2}{\bar{\kappa}}
\int_{-d}^{d} dx'\, K(x,x')\, (n_{0}-n_{{\rm el}}(x')),
\label{eq:VHartree} \ee within the Thomas-Fermi approximation
(TFA), which assumes that the electrostatic quantities vary slowly
in the quantum mechanical scales, such as the magnetic length
$l=\sqrt{\hbar/m \omega_c}$ ($\omega_c=eB/mc$). Eq. (\ref{tfaed})
describes  $n_{\rm el}(x)$ as a function of the electrochemical
potential $\mu^\star$ (which is constant in the absence of $I$),
temperature $(T)$ and total potential energy $V(x)$, where $D(E)$
is the Gaussian broadened density of states (DOS) given by \be
D(E)=\frac{1}{2\pi
l^2}\sum_{n=0}^{\infty}\frac{\exp(-[E_n-E]^2/\Gamma^2)}{\sqrt{\pi}\,\Gamma}
\ee with the impurity parameter $\Gamma$, which gives the Landau
level (LL) broadening and the Landau energy
$E_n=\hbar\omega_c(n+1/2)$. Whereas Eq. (\ref{eq:VHartree})
relates the charge distribution with the total potential. We keep
the donor distribution fixed, with a constant surface number
density $n_0$, and obtain $n_{\rm el}(x)$ iteratively. Here
$\bar{\kappa}$ is an average dielectric constant and $K(x,x')$ is
the solution of the Poisson equation preserving the boundary
conditions, $V(-d)=V(d)=0$. The confinement potential can be
calculated analytically yielding, $V_{\rm
conf}(x)=E_0\sqrt{1-(x/d)^2}$, where $E_0=\frac{2\pi
e^2}{\bar{\kappa}}n_0d$ is the pinch-off energy. In thermal
equilibrium we solve these two equations iteratively, keeping the
average electron density constant, starting from $T=0$ and $B=0$
solutions.

In the presence of an external (fixed) $I$ driven in longitudinal
direction, the situation is fairly different. Since the imposed
current induce modifications on $n_{\rm el}(x)$ and $V(x)$, one
should include this effect self-consistently into the above
scheme, which is done by assuming a local thermal equilibrium. The
driving electric field is given by the gradient of the (now,
position-dependent) electrochemical potential,
$E(\textbf{r})=\nabla \mu^\star(\textbf{r})/e
=\hat{\rho}(\textbf{r})j(\textbf{r})$. With translational
invariance and keeping the intensity of $I$ fixed, one can obtain
the current distribution and the position dependent
electrochemical potential, for a given local resistivity
($\hat{\rho}(\textbf{r})=[\sigma(n_{\rm el}(x))]^{-1}$) tensor. In
the next step this $\mu^\star(x)$ will be used to obtain the new
$n_{\rm el}(x)$ and $V(x)$ in a second iteration loop.

\begin{figure}
{\centering
\includegraphics[width=1.\linewidth]{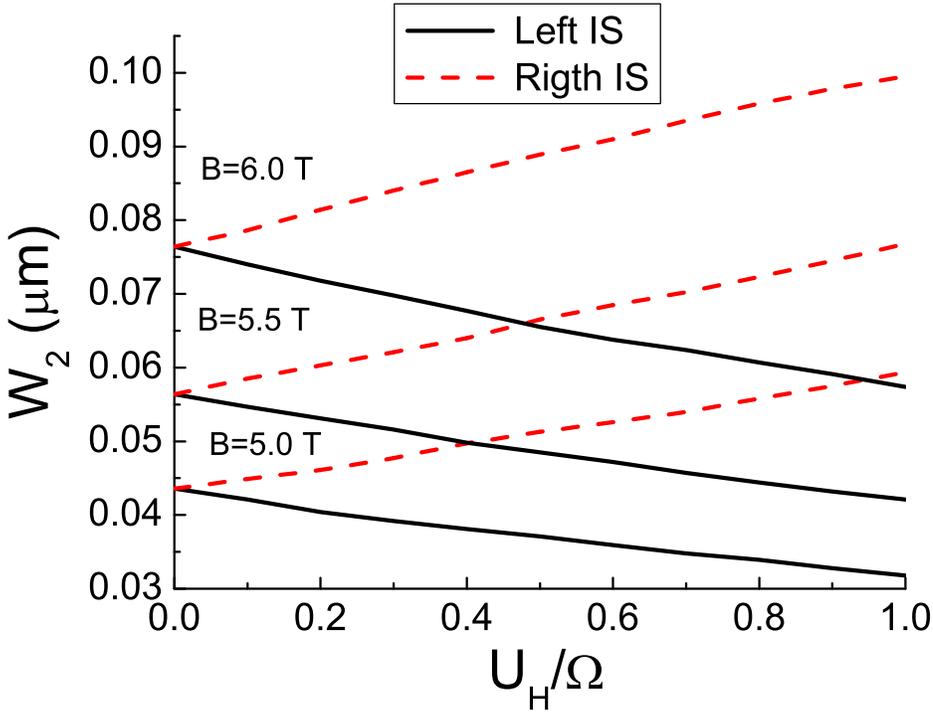}
%
\caption{ \label{fig:fig2} The widths of the ISs at two edges of a
narrow sample ($2d=2 \mu$m) for three typical $B$ values at $T=1
K$ vs. $I$. The widths of the ISs increase (almost) linearly at
the right edge (broken lines), whereas decreases similarly on the
opposite edge. The impurity concentration is chosen such that the
corresponding DOS broadening $\gamma=\Gamma/\hbar \omega_c=0.025$.
The number density of the donors is fixed $n_0=4\cdot10^{11}$
cm$^{-2}$, whereas the depletion length is $200$ nm.}}
\end{figure}

\begin{figure}
{\centering
\includegraphics[width=1.\linewidth]{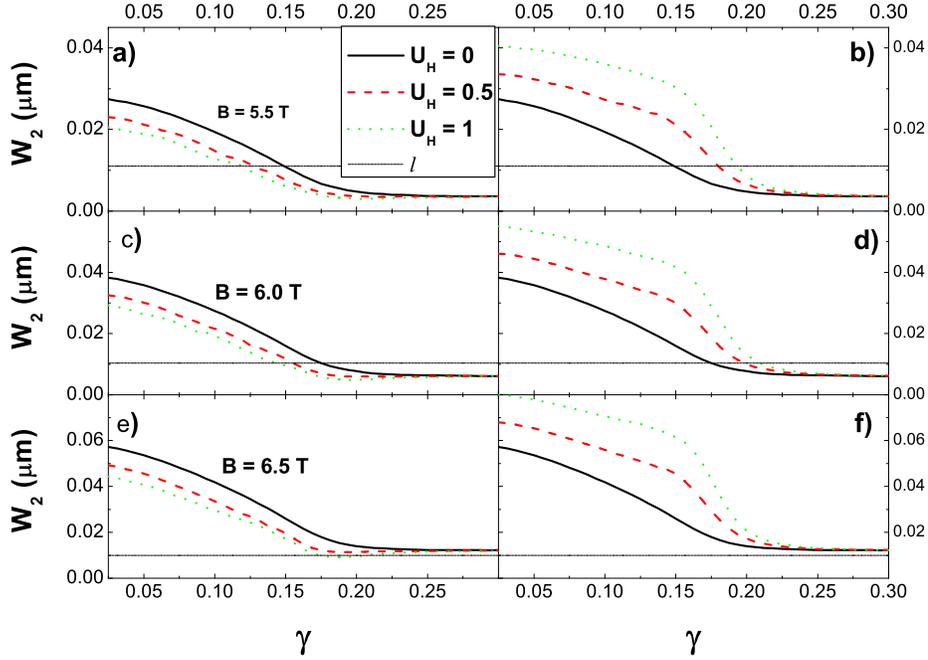}
%
\caption{ \label{fig:fig3} $W_2$ at left (left panel) and right
(right panel) side of the sample as a function of impurity
concentration. At default $T$ and depletion length. Horizontal
(dashed-dotted) lines indicate the magnetic length. The sample
width is taken to be $3$ $\mu$m and calculations are done at $3$
K.}}
\end{figure}

We first investigate the effect of the current intensity
$I=(U_H/e)[d/b](e^2\bar{n}_{\rm el}/m\omega_c)E_0$, measured in
units of $\hbar\omega_c (=\Omega)$ on the widths ($W_2$) of the
ISs with $\nu(x)=2$. For a detailed discussion of the electron and
potential distribution, we suggest the reader to check Fig.1 and
the related text of Ref.~\cite{Guven03:115327}. In Fig.
\ref{fig:fig2}, we show the evolution of the IS widths as a
function of the current intensity for three $B$ values. We see
that $W_2$ on both sides are (almost) linearly dependent on $I$,
for the right IS it increases and for the left IS decreases. We
observe that the total width of the IS increases by increasing the
intensity.
For a given magnetic field, the average electron velocity is
defined as $ v_{\rm {y}}(x)=\frac{1}{\hbar}\frac{\partial
V(x)}{\partial x}$,  within the TFA imposing $E_n(X)=E_n+V(X)$ and
the center coordinate can be replaced by the spatial coordinate
$x$. If one calculates the $v_{\rm el}$ at the right IS, it will
decrease by increasing $I$, since at a fixed $B$, the hight of the
potential drop remains constant, namely $\Omega$, meanwhile the
thickness of the IS increases. We should also note that, although
the velocity of the electrons decrease, the number of them
increase due to a wider IS, therefore more current is carried at
the right IS over all. In Fig.~\ref{fig:fig3}, we plot the effect
of DOS broadening on the IS widths, again for three selected $B$
field values and current intensities. It is known that if the
widths of the ISs become small or comparable with the magnetic
length they essentially
disappear~\cite{siddiki2004,Suzuki93:2986}, however, here we still
observe them as an artifact of TFA. We see that the left ISs
become narrower than the magnetic length compared to the right
ISs, due to the strong current-induced-asymmetry. We observe that
the impurity concentration effects the $W_2$ in a non-linear
manner strongly and if the DOS broadening exceeds $\% 20$ of the
cyclotron energy, no ISs are left. At zero bias the transition
between having an IS or not is rather smooth, meanwhile this
transition occurs much drastically in the case of $I\neq0$.

To summarize, we have studied the widths of the ISs, in the OLR
regime. We found that (i) the strong current imposed, induces an
asymmetry on the IS width depending linearly on the current
intensity; (ii) the higher the impurity concentration, the
narrower the $W_2$ is, meanwhile at higher currents this effect
becomes more pronounced. The main message of our self-consistent
calculations is that the electron velocity strongly depends on the
sample parameters and, in addition, in the OLR regime the symmetry
between the left and right edges is broken due to
electron-electron interaction.

The authors acknowledge the support of the Marmaris Institute of
Theoretical and Applied Physics (ITAP), TUBITAK grant $105T110$,
TUBAP-739-754-759, SFB631 and DIP.


\begin{thebibliography}{10}

\bibitem{Ahlswede02:165}
E. Ahlswede, J. Weis, K. von Klitzing and K. Eberl, Physica E {\bf
12},  165
  (2002).

\bibitem{Yacoby04:328}
S. Ilani, J. Martin, E. Teitelbaum, J.~H. Smet, D. Mahalu, V.
Umansky and A.
  Yacoby, Nature {\bf 427},  328  (2004).

\bibitem{Guven03:115327}
K. G{\"u}ven and R.~R. Gerhardts, Phys. Rev. B {\bf 67},  115327
(2003).

\bibitem{siddiki2004}
A. Siddiki and R.~R. Gerhardts, Phys. Rev. B {\bf 70},  195335
(2004).

\bibitem{TobiasK06:h}
T. {Kramer}, International Journal of Modern Physics B {\bf 20},
1243  (2006).

\bibitem{Romer07}
C. {Sohrmann} and R.~A. {R{\"o}mer}, New Journal of Physics {\bf
9},  97
  (2007).

\bibitem{Neder06:016804}
I. Neder, M. Heiblum, Y. Levinson, D. Mahalu and V. Umansky, Phys.
Rev. Lett.
  {\bf 96},  016804  (2006).

\bibitem{deniz06}
D. {Eksi}, E. {Cicek}, A.~I. {Mese}, S. {Aktas}, A. {Siddiki} and
T.
  {Hakioglu}, cond-mat/0612519  (2006).

\bibitem{Chklovskii93:12605}
D.~B. Chklovskii, K.~A. Matveev and B.~I. Shklovskii, Phys. Rev. B
{\bf 47},
  12605  (1993).

\bibitem{Suzuki93:2986}
T. Suzuki and T. Ando, J. Phys. Soc. Jpn. {\bf 62},  2986  (1993).

\end{thebibliography}

\end{document}